\documentclass[
    ,final            
  ]
  {aipproc}

\layoutstyle{6x9}

\begin{document}

\title{The nature of the soft X-ray emission in obscured AGN}


\author{Stefano Bianchi}{
  address={Dipartimento di Fisica, Universit\`a degli Studi Roma Tre, Italy}
   ,altaddress={XMM-Newton Science Operations Center, European Space Astronomy Center, ESA, Apartado 50727, E-28080 Madrid, Spain}
}

\author{Matteo Guainazzi}{
  address={XMM-Newton Science Operations Center, European Space Astronomy Center, ESA, Apartado 50727, E-28080 Madrid, Spain}
}

\begin{abstract}
The origin of the soft X-ray emission in obscured AGN is still largely unknown. However, important progresses have been made thanks to the high energy and spatial resolution of XMM-\textit{Newton} and \textit{Chandra}. We review here the latest results on this issue, focusing on the physical properties of the material responsible for the soft X-ray emission and its relation to the circumnuclear environment, putting them in the general context of our understanding of the AGN structure and its feedback to the host galaxy.
\end{abstract}

\maketitle


\section{Introduction}

The spectra of nearby X-ray obscured Active Galactic Nuclei (AGN) invariably present a `soft excess', i.e.  soft X-ray emission above the extrapolation of the absorbed nuclear emission \citep{gua05b,turner97}. Low resolution, CCD data are generally insufficient to discriminate between thermal emission, as expected by gas heated by shocks induced by AGN outflows or episodes of intense star formation, and emission from a gas photoionized by the AGN primary emission. However, important progresses have been made in the last few years, thanks to the high energy and spatial resolution of XMM-\textit{Newton} and \textit{Chandra}. Understanding the nature of this component has profound implications not only for the general unification models of AGN, but also for the relationships between the latter and the host galaxy. 

\section{High Resolution Spectroscopy: the key evidence}

The first breakthrough was represented by high resolution spectra made available thanks to the gratings aboard \textit{Chandra} and XMM-\textit{Newton}. The observations of a few bright objects revealed that the `soft excess' observed in CCD spectra was due to the blending of bright emission lines, mainly from He- and H-like transitions of light metals and L transitions of Fe, with low or no continuum \citep[e.g.][]{sako00b,Sambruna01b,kin02,brink02,schurch04,bianchi05b,pp05}. Narrow Radiative Recombination Continua (RRC) features from Carbon and Oxygen were detected, whose width indicates typical plasma temperatures of the order of a few eV \citep{kin02}. These features are unequivocal signatures of photoionized spectra \citep{lp96}. Moreover, the intensity of higher-order series emission lines, once normalized to the K$_{\alpha}$, are larger than predicted
by pure photoionization, and are consistent with an important contribution by photoexcitation (resonant scattering) \citep{band90,matt94,kk95}. This explains why standard plasma diagnostics \citep{pd00} fail to properly interpret the physical nature of the spectra. All these pieces of evidence agree that the observed lines should be produced in a gas photoionized by the AGN, with little contribution from any collisionally ionized plasma.

\begin{figure}
\includegraphics[height=.5\textwidth]{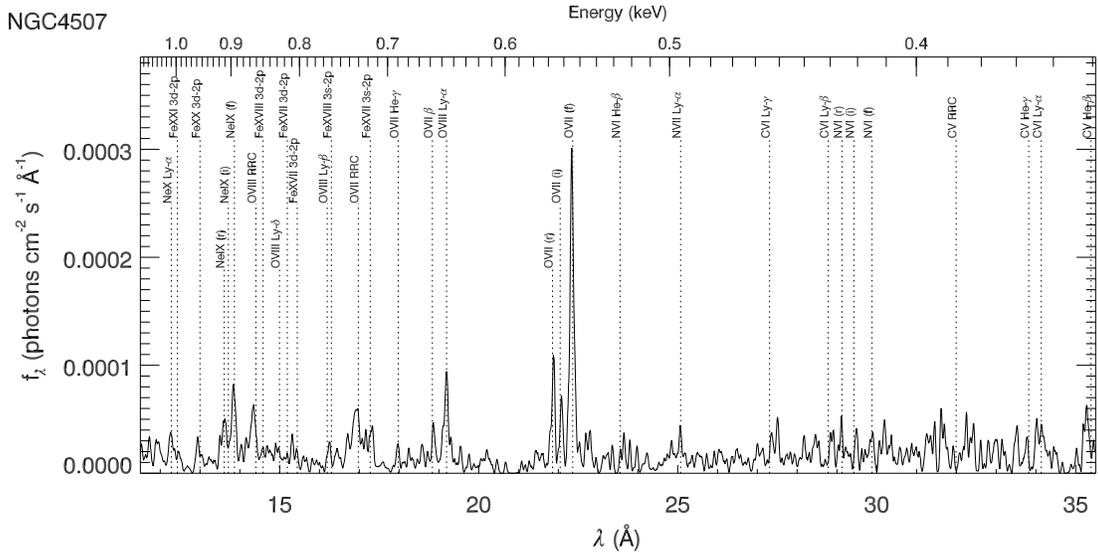}
\caption{\label{rgs1}RGS spectrum of NGC~4507, one of the Seyfert 2 galaxies included in CIELO. Spectra of the two RGS cameras have been merged and smoothed with a 5-channels wide triangular kernel for illustration purposes only. From \citet{gb06}. }
\end{figure}
\begin{figure}
\includegraphics[height=.5\textwidth]{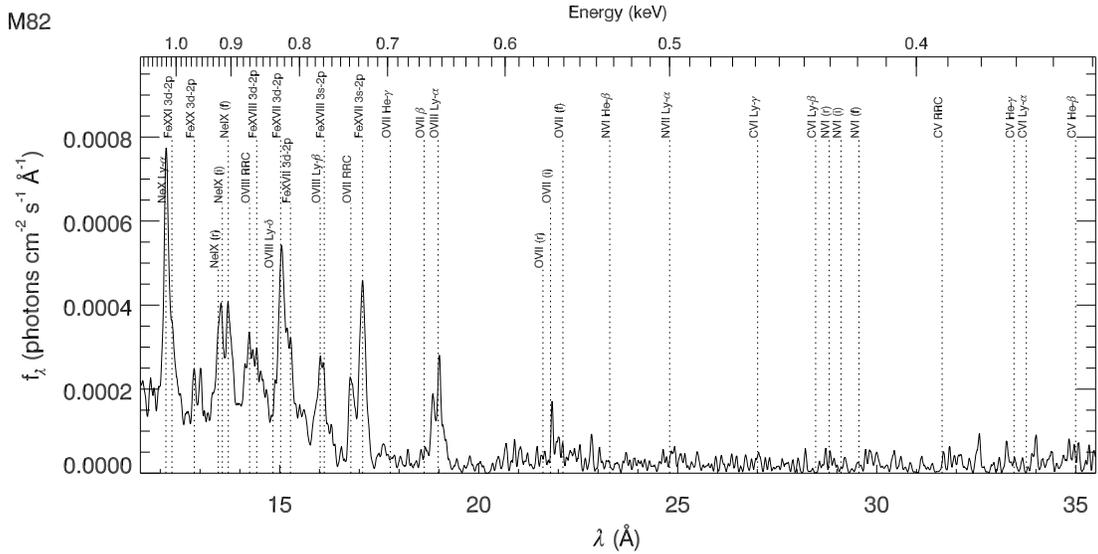}
\caption{\label{rgs2}Same as Figure \ref{rgs1}, but for M~82, one of the starburst galaxies included in the control sample of CIELO.}
\end{figure}

\begin{figure}
\includegraphics[height=.33\textheight,width=.53\textwidth]{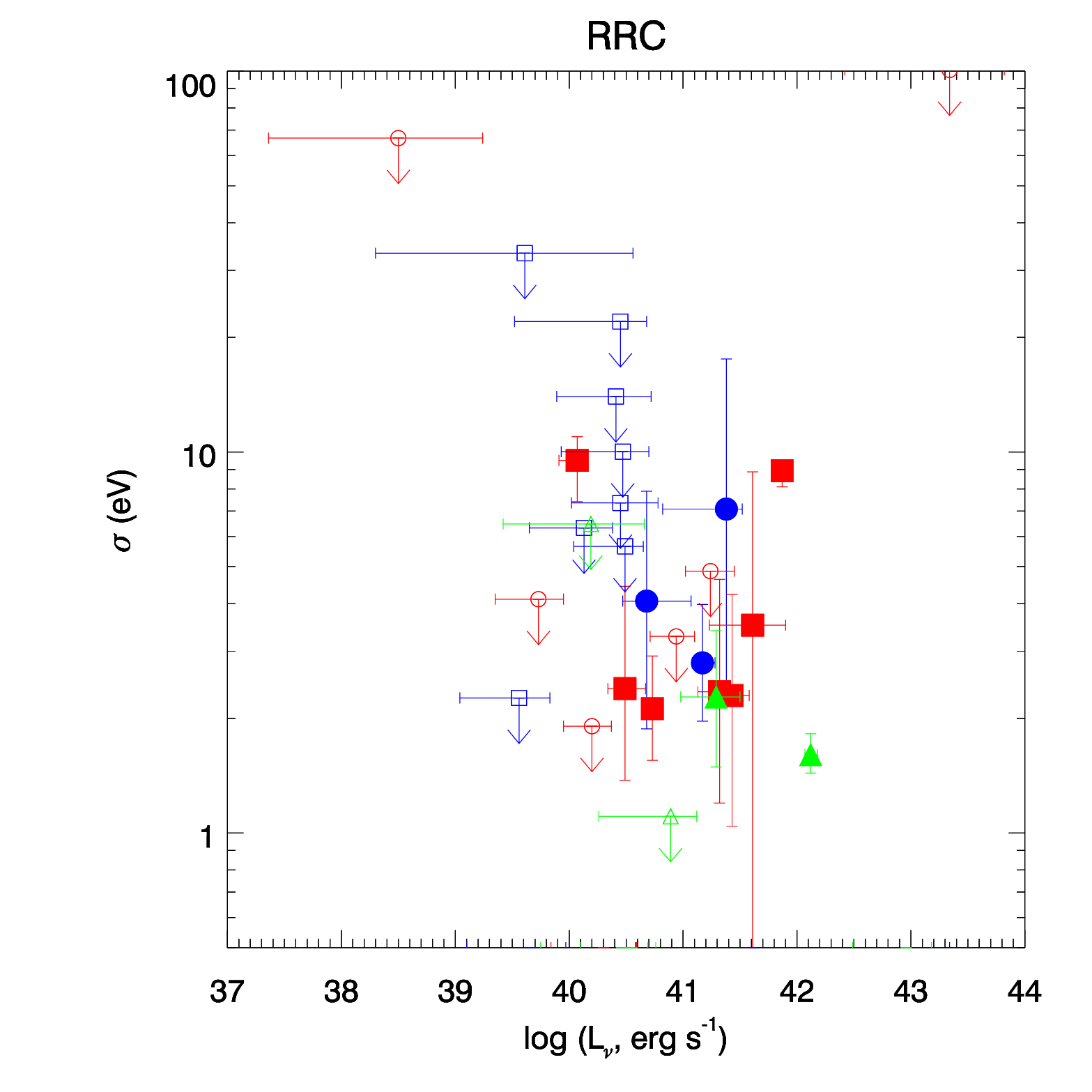}
\includegraphics[height=.33\textheight,width=.53\textwidth]{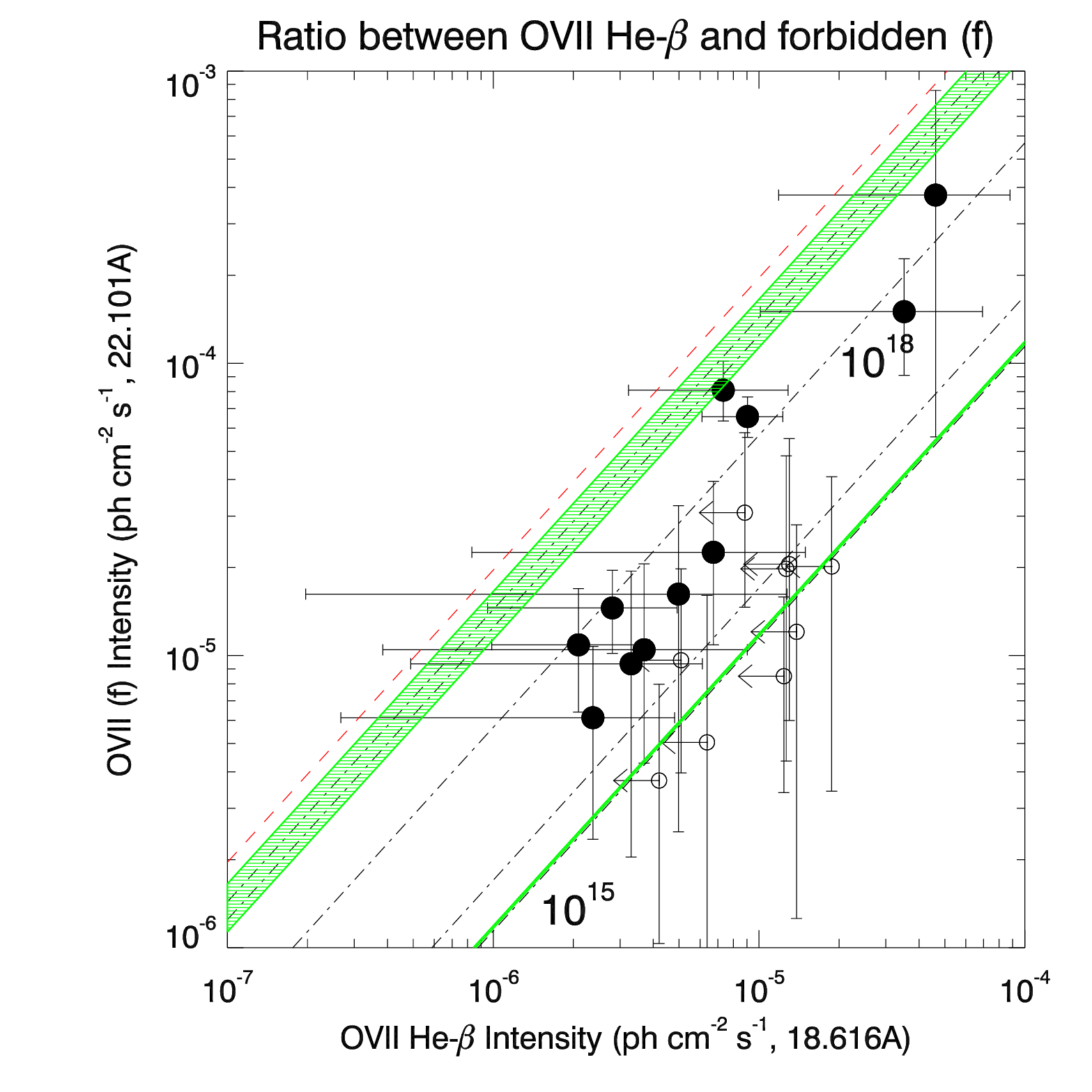}
\caption{\label{cielo}Left: Luminosity versus intrinsic width for the RRC features detected in CIELO \citep{gb06}. {\it Filled} data points correspond to a
measurement of the RRC width; {\it empty} data points correspond to RRC width upper limits. {\it Circles}: O{\sc viii}; {\it squares}: O{\sc vii};
{\it triangles}: C{\sc v}. Data points corresponding to upper limits on both quantities are not shown for clarity. Right: Intensity of O{\sc vii} He-$\beta$
line against the intensity of the $f$ component of the He-$\alpha$ triplet (only data points corresponding to a detection of the latter are shown; data points correspond to upper limits on the intensity of the former are shown as {\it empty symbols}). The {\it dashed-dotted lines} represent the prediction
of the {\tt photoion} code for O{\sc vii} column densities increasing from $N_{OVII} = 10^{15}$ to $10^{20}$~cm$^{-2}$ in steps of one decade, assuming $kT = 5$~eV and $v_{turb} = 200$~km~s$^{-1}$. The {\it long-dashed line} represents the predictions for pure photoionization. See \citep{gb06} for details.}
\end{figure}

\begin{figure}
\includegraphics[height=.45\textheight]{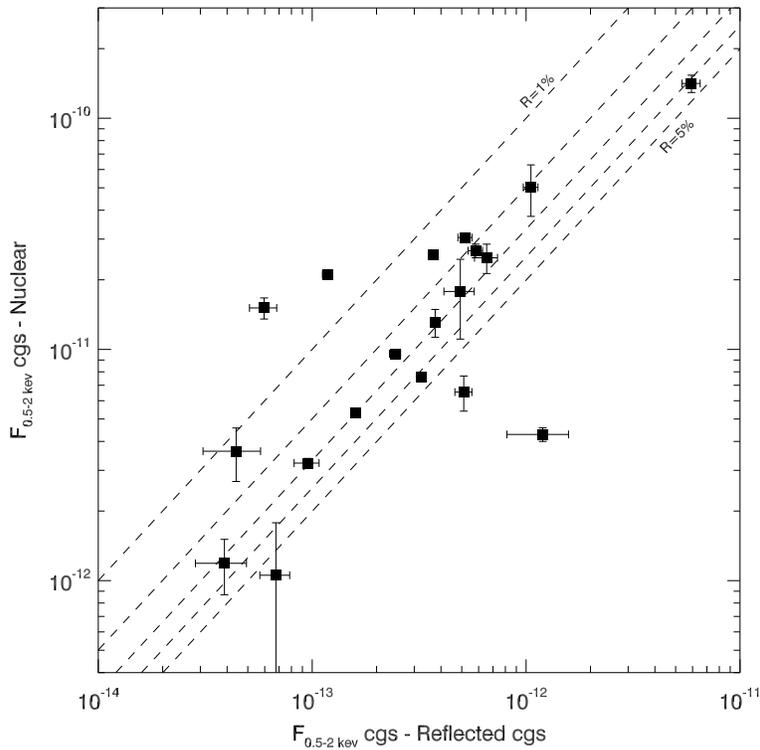}
\caption{\label{sy2R}Reflected soft X-ray flux against extrapolated nuclear soft X-ray flux for a sample of unobscured ($10^{22}<N_{H}<10^{24}\,cm^{-2}$) AGN observed by XMM-\textit{Newton}. The dotted lines represent values of 1 to 5 \% for that ratio. From Bianchi et al., in preparation.}
\end{figure}

These results have been recently confirmed to be common in a large catalog of Seyfert 2 galaxies (Catalog of Ionized Emission Lines in Obscured AGN: CIELO-AGN \citet{gb06}), who presented results of a high-resolution soft X-ray (0.2--2~keV) spectroscopic study on a sample of 69 nearby obscured Active Galactic Nuclei (AGN) observed by the Reflection Grating Spectrometer (RGS) on board XMM-Newton. This study allowed them to explore the nature of the soft X-ray emission in objects as weak as $\sim$0.03~mCrab (see Figure \ref{rgs1} for the spectrum of an object of their catalog). Their analysis confirmed the dominance of emission lines over the continuum in the soft X-ray band of these sources, the presence of narrow RRC and the important contribution from higher-order series lines (see Figure \ref{cielo}). Therefore, this study allows us to confirm that the results extracted from the detailed study of high-quality spectra of the brightest objects can be extended to the whole population of nearby obscured AGN. In agreement with this scenario, where the soft X-ray emission in these objects is due to reprocessing of the primary emission, the observed soft X-ray emission and the nuclear, unabsorbed soft X-ray emission are well correlated, and their ratio is around 3\% (see Figure \ref{sy2R}).


\section{Location of the gas: a single photoionized medium?}

\begin{figure}
\includegraphics[height=.34\textheight]{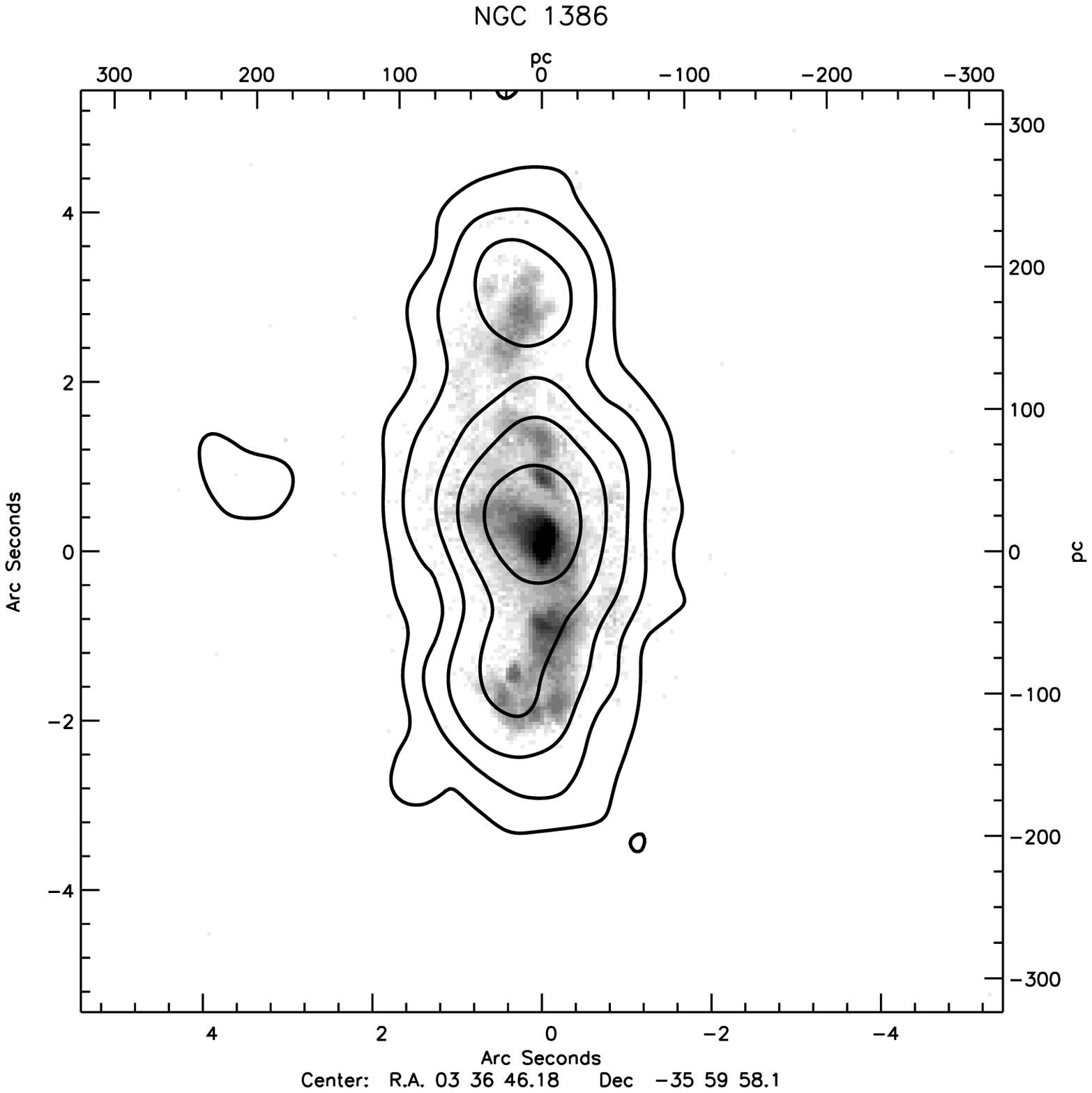}
\includegraphics[height=.34\textheight]{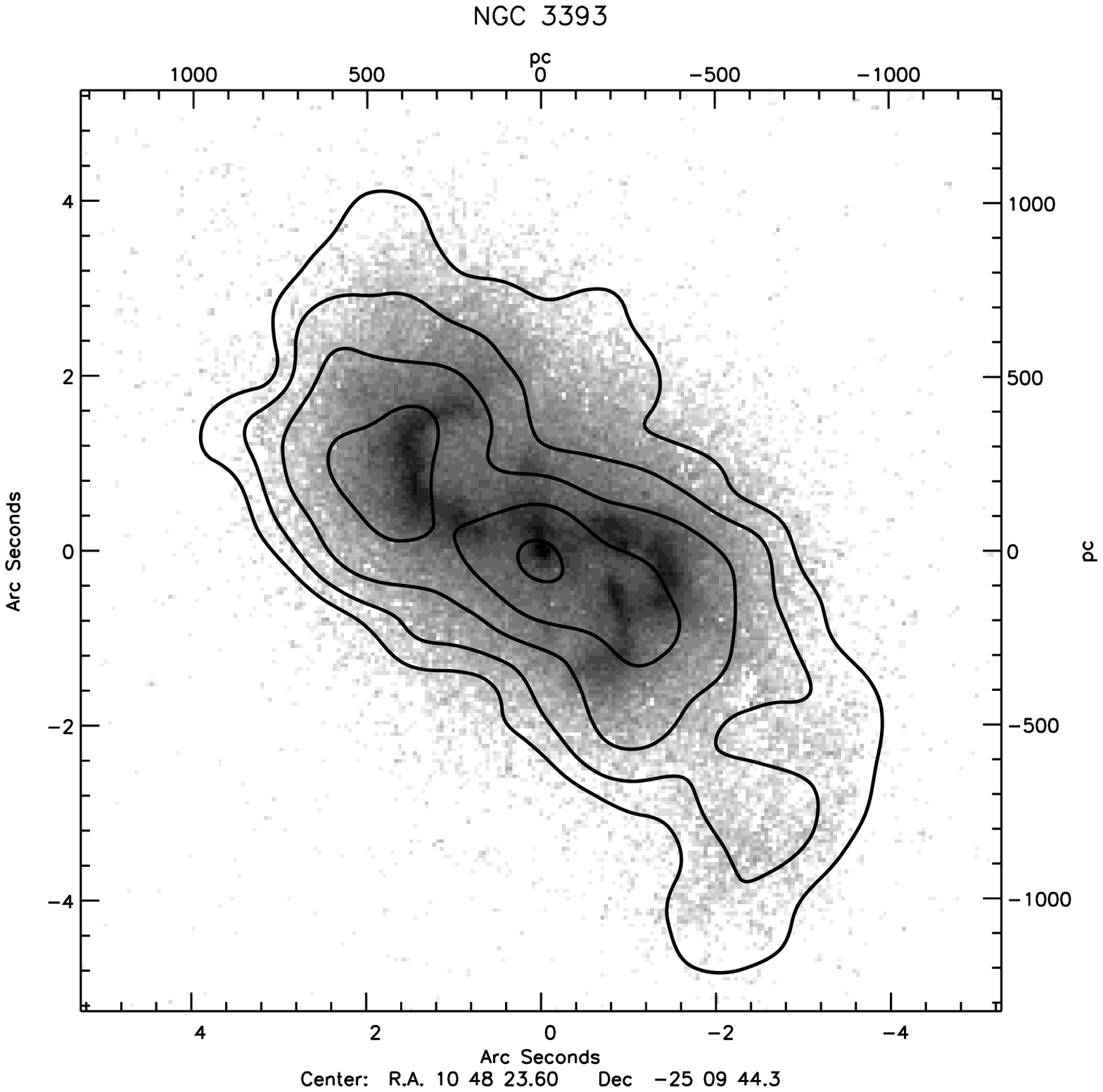}
\caption{\label{xray2oiiimaps}\textit{Chandra} soft X-ray contours superimposed on \textit{HST} [{O\,\textsc{iii}}] images. The contours correspond to five logarithmic intervals in the range of 1.5-50\% (NGC~1386) and 5-90\% (NGC~3393). The \textit{HST} images are scaled with the same criterion for each source. From \citet{bianchi06}}
\end{figure}

A second breakthrough in understanding the nature of the soft X-ray emission in obscured AGN was made possible thanks to the unrivaled spatial resolution of \textit{Chandra}. The soft X-ray emission of Seyfert 2 galaxies appears to be strongly correlated with that of the Narrow Line Region (NLR), as mapped by the [{O\,\textsc{iii}}] $\lambda 5007$ \textit{HST} images \citep[e.g.][see Figure \ref{xray2oiiimaps}]{yws01,bianchi06,levenson06}.

\begin{figure}
\includegraphics[height=.3\textheight,width=.5\textwidth]{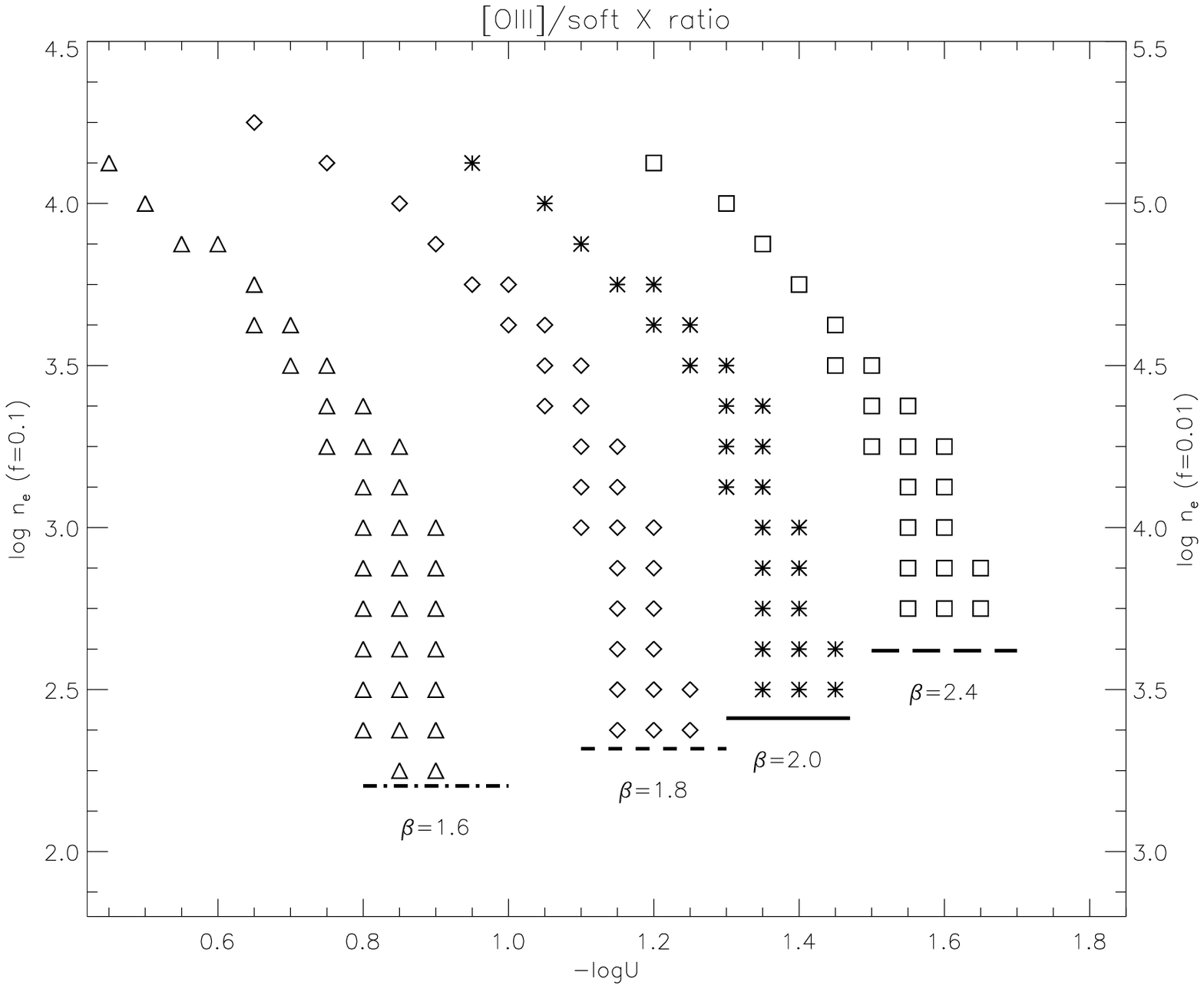}
\includegraphics[height=.3\textheight,width=.49\textwidth]{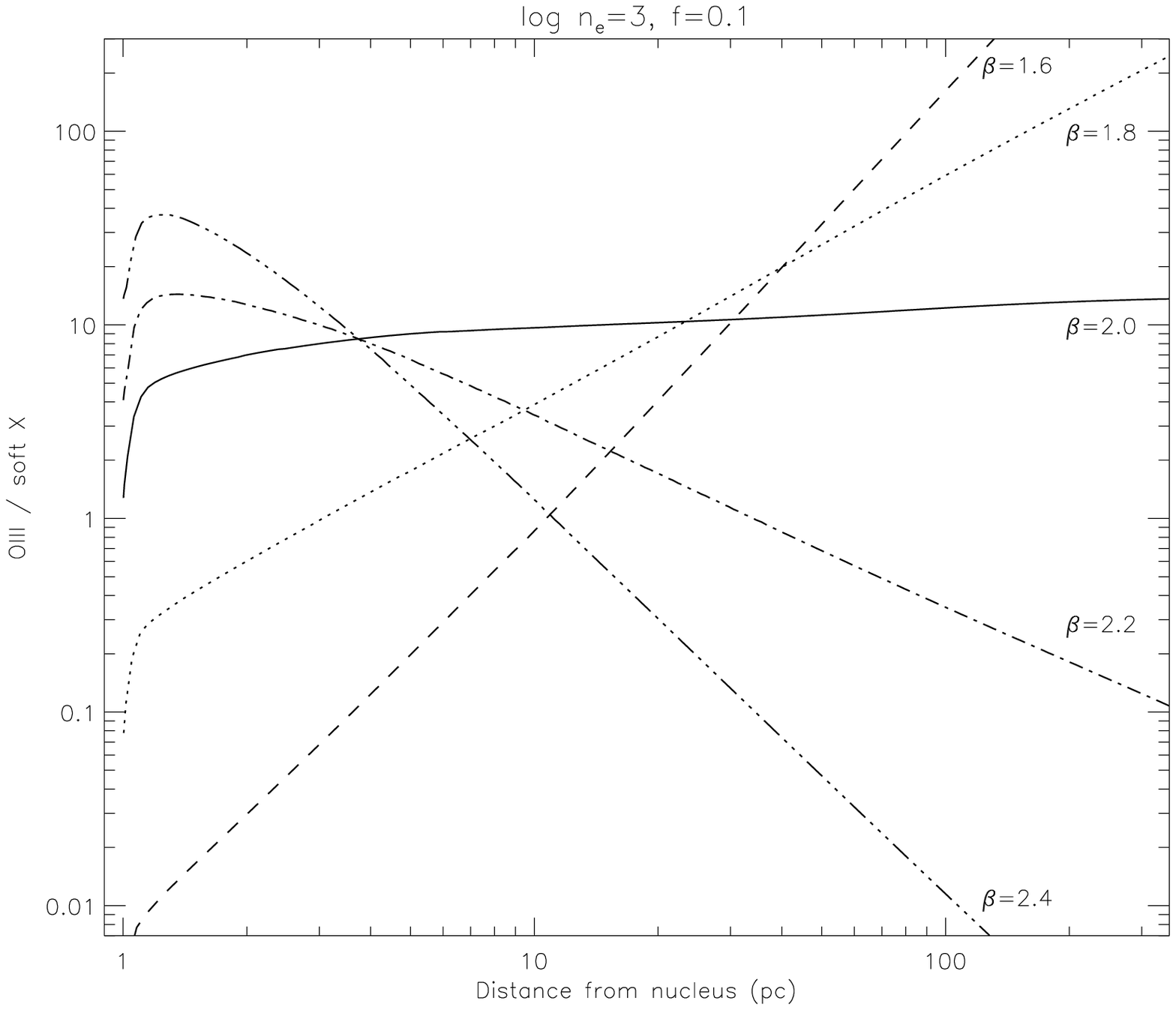}
\caption{\label{cone_ratio_r}Left: Each symbol represents one solution in our grid of \textsc{Cloudy} models that satisfies the condition of total [{O\,\textsc{iii}}] to soft X-ray flux ratio in the range observed in our sample, plotted in a three-parameter space $U$, $n_e$ and $\beta$, i.e. the ionization parameter and the density at the beginning of the cone of gas (1 pc), and the index of the density powerlaw, represented by different symbols (\textit{triangles}: $\beta=1.6$, \textit{diamonds}: $\beta=1.8$, \textit{stars}: $\beta=2.0$, \textit{squares}: $\beta=2.4$). The horizontal lines determine the limit corresponding to a total column density of $10^{20}$ cm$^{-2}$ for each index: solutions below this limit are not plotted. Solutions with $\beta=2.2$ are not plotted for clarity reasons. The net effect of the filling factor $f$ is to shift the density values: the two y axes refer to $f=0.1$ and $f=0.01$ (see text for details). Right: The [{O\,\textsc{iii}}] to soft X-ray ratio plotted as a function of the radius of the gas, for different values of $\beta$, when $\log{n_e}=3$ and $f=0.1$. The corresponding values of $\log{U}$ for these solutions are (from top to bottom): -0.85, -1.15, -1.4, -1.5, -1.6. From \citet{bianchi06}.}
\end{figure}

The possibility that the NLR, which is also believed to be gas photoionized by the AGN, is the same material responsible for the soft X-ray emission was investigated in detail by \citet{bianchi06}. They found that such a simple scenario is tenable.  Moreover, the observed [{O\,\textsc{iii}}] to soft X-ray flux ratio remains fairly constant up to large radii, thus requiring that the density decreases roughly like $r^{-2}$, similarly to what often found for the Narrow Line Region \citep[e.g.][]{krae00b,coll05} (see Figure \ref{cone_ratio_r}).

In any case, these models are clearly oversimplified and do not take into account a number of details. In particular, there is evidence that various components are present at any radius, as also indicated by the clumpiness of the extended region when seen with the high spatial resolution of \textit{HST}. The soft X-ray emitting region also appears dishomogeneous, when seen with enough spatial resolution and statistics. In NGC~7582, for example, two `hot spots' where detected, i.e. regions where the ionization parameter is likely to be larger than the average \citep{bianchi06b}. In this sense, the role of radio jets, almost always found to be very well correlated with the NLR/soft X-ray emission, is still under debate and could be very important at least for the morphology of the gas.

\section{Starbursts}

It is very difficult to disentangle the starburst contribution from the total soft X-ray emission of Seyfert 2s. In some cases, the correlation of the optical, IR and X-ray images strongly suggests that the star-formation regions cannot account for a significant part of the extended X-ray emission \citep{yws01,bianchi06b}. On the other hand, there are also clear examples where the starburst contribution appears to be at least comparable with that from the AGN \citep{jb03,jb05b}. And there are cases where such a contribution seems to be dominant, on the basis of the total Spectral Energy Distribution (SED) of the object \citep{levenson05}. 

\begin{figure}
\includegraphics[height=.35\textheight]{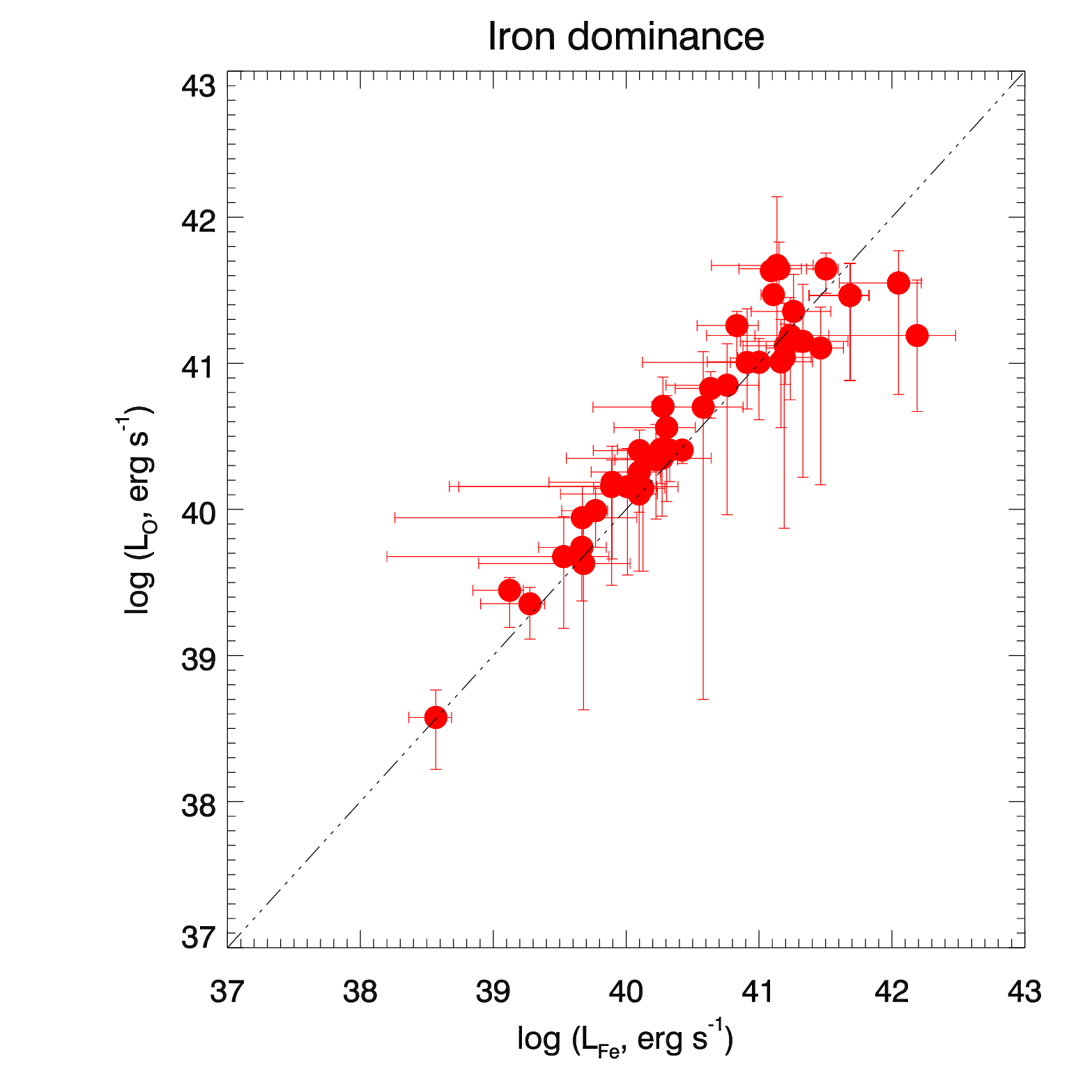}
\includegraphics[height=.35\textheight]{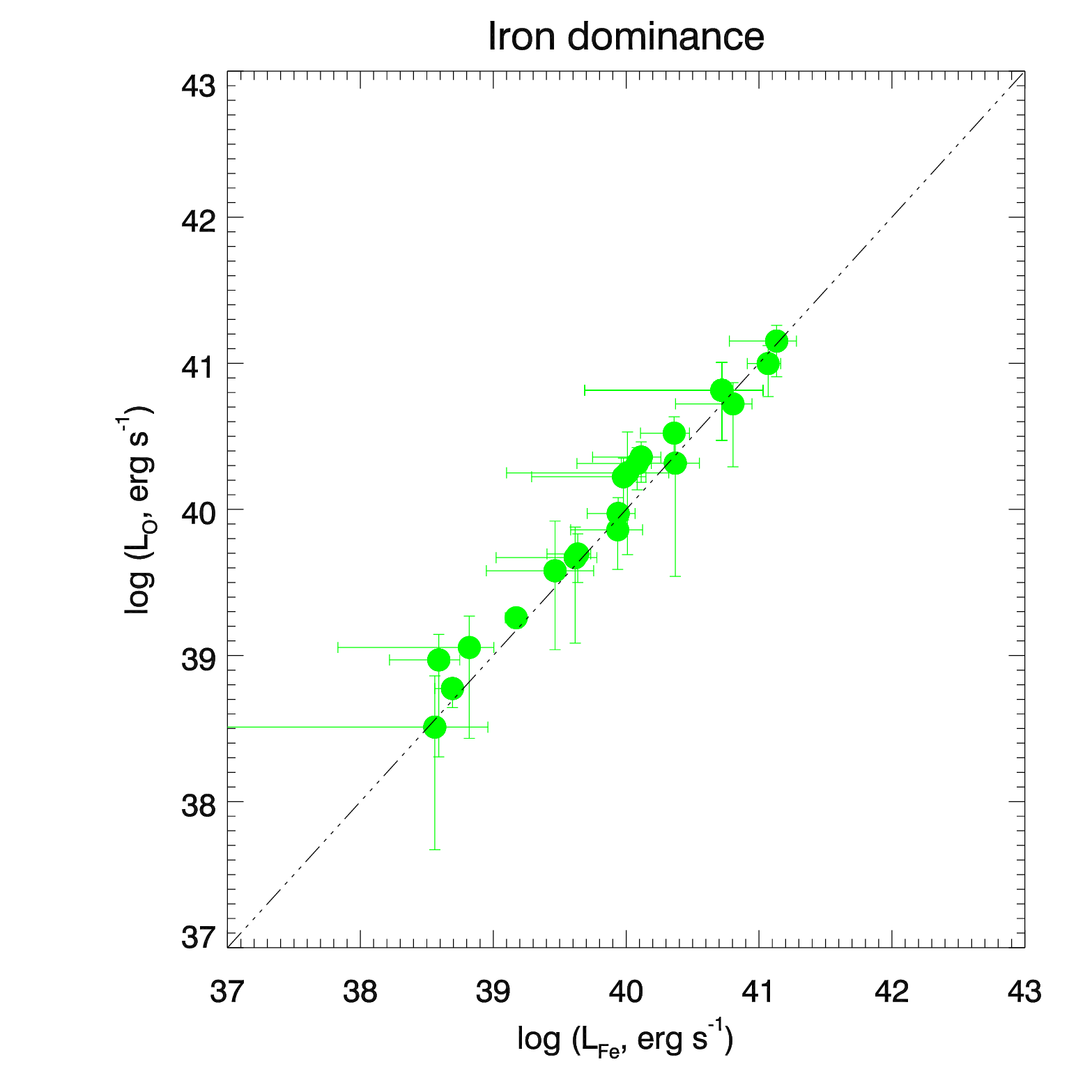}
\caption{\label{irondom}Ratio between the integrated flux of iron L emission lines and of oxygen K lines for the Seyferts 2s and starburst galaxies included in CIELO \citep{gb06}. No difference is apparent between the two populations. In particular, there is no evidence of an iron dominance in starburst galaxies.}
\end{figure}

\begin{figure}
\includegraphics[height=.45\textheight]{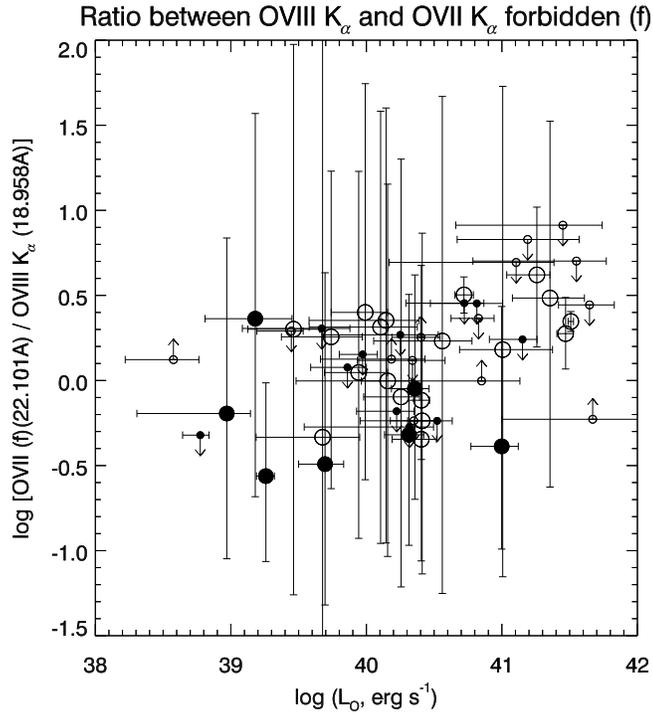}
\caption{\label{starburst}Intensity of the $f$
component of the O{\sc vii} triplet (normalized to the O{\sc viii} Ly-${\alpha}$ intensity) against the total luminosity in Oxygen lines.
{\it Empty circles} represents the obscured AGN in {\it CIELO-AGN}, {\it filled circles} the control sample of 27 starburst galaxies.
Symbols representing censored data are plotted with a smaller size for the sake of clarity. From \citet{gb06}}
\end{figure}

In their analysis of RGS spectra of Seyfert 2s, \citet{gb06} studied a control sample of Starburst galaxies, in order to find a way to discriminate between the two via high resolution spectroscopy. Although in some cases the spectra of the two classes of objects are clearly different (see Figure \ref{rgs1} and \ref{rgs2}), some of the discriminating criteria quoted in the literature are inconclusive. For example, there is no evidence of an iron L dominance over oxygen H- and He-like emission lines (Figure \ref{irondom}). A new criterion is shown in Figure \ref{starburst}, where the distribution of the O{\sc vii} $f$ transition intensity, normalized against the intensity of the O{\sc viii} Ly-${\alpha}$ is shown as a function of the total luminosity in oxygen lines (integrated on all He- and H-like transitions). AGN are generally characterized by larger {O\,\textsc{viii}}/{O\,\textsc{vii}} ratios. The same figure as well shows that the average line luminosity in AGN is larger than in starbursts. Still, the overlap between the line diagnostic distributions is significant, and prevents strong statements on individual sources. This most likely reflects the intrinsic ``composite'' nature of several obscured AGN \cite{cidfernandes01}. 



\section{The big picture}

High resolution spectroscopy of soft X-ray emission in obscured AGN reveals that it is dominated by strong emission lines, with a low level of continuum. Several diagnostic tools both on few bright sources and on a large sample (CIELO) agree on its origin in a photoionized gas, with photoexcitation playing an important role. The main properties of this gas (column density, ionization parameter) are very similar to those found for the warm absorbers observed in type 1 objects, suggesting a coincidence between the two media, with the observing line of sight playing the key role.

The common spatial coincidence between the soft X-ray emission and the NLR suggests they can also be one and the same medium, photoionized by the central AGN, as confirmed by simple models. These models require that the gas density should decrease with distance roughly like $r^{-2}$, suggestive of outflow emission. This is again in agreement with current observational evidence about warm absorbers \citep[see e.g.][]{blust05} and with modified unification models, such the one presented by \citet{elvis00}. 

The presence of an outflow extending from the nucleus up to distances of the orders of kpc, as suggested by these results, has important consequences on the feedback between the AGN and the host galaxy. In particular, it can be a powerful way to provide metals to the galaxy, since the metallicity of the nuclear regions is believed to be much higher than solar \citep[see e.g.][]{nagao06}. Finally, the difficulties in discriminating between AGN and starburst emission and the presence of many composite objects strongly suggest that they may be related one to the other. This should be taken into account whenever one tries to understand the two phenomena.



\bibliographystyle{aipproc}   

\bibliography{sbs}

\begin{thebibliography}{28}
\expandafter\ifx\csname natexlab\endcsname\relax\def\natexlab#1{#1}\fi
\providecommand{\enquote}[1]{``#1''}
\expandafter\ifx\csname url\endcsname\relax
  \def\url#1{\texttt{#1}}\fi
\expandafter\ifx\csname urlprefix\endcsname\relax\def\urlprefix{URL }\fi
\providecommand{\eprint}[2][]{\url{#2}}

\bibitem[{Guainazzi} et~al.(2005)]{gua05b}
M.~{Guainazzi}, G.~{Matt}, and G.~C. {Perola}, \emph{\aap} \textbf{444},
  119--132 (2005).

\bibitem[{Turner} et~al.(1997)]{turner97}
T.~J. {Turner}, I.~M. {George}, K.~{Nandra}, and R.~F. {Mushotzky},
  \emph{\apjs} \textbf{113}, 23--+ (1997).

\bibitem[{Sako} et~al.(2000)]{sako00b}
M.~{Sako}, S.~M. {Kahn}, F.~{Paerels}, and D.~A. {Liedahl}, \emph{\apjl}
  \textbf{543}, L115--L118 (2000).

\bibitem[Sambruna et~al.(2001)]{Sambruna01b}
R.~M. Sambruna, H.~Netzer, S.~Kaspi, W.~N. Brandt, G.~Chartas, G.~P. Garmire,
  J.~A. Nousek, and K.~A. Weaver, \emph{\apj} \textbf{546}, L13 (2001).

\bibitem[{Kinkhabwala} et~al.(2002)]{kin02}
A.~{Kinkhabwala}, M.~{Sako}, E.~{Behar}, S.~M. {Kahn}, F.~{Paerels}, A.~C.
  {Brinkman}, J.~S. {Kaastra}, M.~F. {Gu}, and D.~A. {Liedahl}, \emph{\apj}
  \textbf{575}, 732--746 (2002).

\bibitem[{Brinkman} et~al.(2002)]{brink02}
A.~C. {Brinkman}, J.~S. {Kaastra}, R.~L.~J. {van der Meer}, A.~{Kinkhabwala},
  E.~{Behar}, S.~M. {Kahn}, F.~B.~S. {Paerels}, and M.~{Sako}, \emph{\aap}
  \textbf{396}, 761--772 (2002).

\bibitem[{Schurch} et~al.(2004)]{schurch04}
N.~J. {Schurch}, R.~S. {Warwick}, R.~E. {Griffiths}, and S.~M. {Kahn},
  \emph{\mnras} \textbf{350}, 1--9 (2004).

\bibitem[{Bianchi} et~al.(2005)]{bianchi05b}
S.~{Bianchi}, G.~{Miniutti}, A.~C. {Fabian}, and K.~{Iwasawa}, \emph{\mnras}
  \textbf{360}, 380--389 (2005).

\bibitem[{Pounds} and {Page}(2005)]{pp05}
K.~A. {Pounds}, and K.~L. {Page}, \emph{\mnras} \textbf{360}, 1123--1131
  (2005).

\bibitem[{Liedahl} and {Paerels}(1996)]{lp96}
D.~A. {Liedahl}, and F.~{Paerels}, \emph{\apjl} \textbf{468}, L33 (1996).

\bibitem[{Band} et~al.(1990)]{band90}
D.~L. {Band}, R.~I. {Klein}, J.~I. {Castor}, and J.~K. {Nash}, \emph{\apj}
  \textbf{362}, 90--99 (1990).

\bibitem[{Matt}(1994)]{matt94}
G.~{Matt}, \emph{\mnras} \textbf{267}, L17--L20 (1994).

\bibitem[{Krolik} and {Kriss}(1995)]{kk95}
J.~H. {Krolik}, and G.~A. {Kriss}, \emph{\apj} \textbf{447}, 512--+ (1995).

\bibitem[{Porquet} and {Dubau}(2000)]{pd00}
D.~{Porquet}, and J.~{Dubau}, \emph{\aaps} \textbf{143}, 495--514 (2000).

\bibitem[{Guainazzi} and {Bianchi}(2006)]{gb06}
M.~{Guainazzi}, and S.~{Bianchi}, \emph{\mnras, in press}  (2006),
  \eprint{astro-ph/0610715}.

\bibitem[{Young} et~al.(2001)]{yws01}
A.~J. {Young}, A.~S. {Wilson}, and P.~L. {Shopbell}, \emph{\apj} \textbf{556},
  6--23 (2001).

\bibitem[{Bianchi} et~al.(2006{\natexlab{a}})]{bianchi06}
S.~{Bianchi}, M.~{Guainazzi}, and M.~{Chiaberge}, \emph{\aap} \textbf{448},
  499--511 (2006{\natexlab{a}}).

\bibitem[{Levenson} et~al.(2006)]{levenson06}
N.~A. {Levenson}, T.~M. {Heckman}, J.~H. {Krolik}, K.~A. {Weaver}, and P.~T.
  {{\.Z}ycki}, \emph{\apj} \textbf{648}, 111--127 (2006).

\bibitem[{Kraemer} et~al.(2000)]{krae00b}
S.~B. {Kraemer}, D.~M. {Crenshaw}, J.~B. {Hutchings}, T.~R. {Gull}, M.~E.
  {Kaiser}, C.~H. {Nelson}, and D.~{Weistrop}, \emph{\apj} \textbf{531},
  278--295 (2000).

\bibitem[{Collins} et~al.(2005)]{coll05}
N.~R. {Collins}, S.~B. {Kraemer}, D.~M. {Crenshaw}, J.~{Ruiz}, R.~{Deo}, and
  F.~C. {Bruhweiler}, \emph{\apj} \textbf{619}, 116--133 (2005).

\bibitem[{Bianchi} et~al.(2006{\natexlab{b}})]{bianchi06b}
S.~{Bianchi}, M.~{Chiaberge}, E.~{Piconcelli}, and M.~{Guainazzi}, \emph{MNRAS
  in press}  (2006{\natexlab{b}}), \eprint{astro-ph/0610277}.

\bibitem[{Jim{\'e}nez-Bail{\'o}n} et~al.(2003)]{jb03}
E.~{Jim{\'e}nez-Bail{\'o}n}, M.~{Santos-Lle{\'o}}, J.~M. {Mas-Hesse},
  M.~{Guainazzi}, L.~{Colina}, M.~{Cervi{\~n}o}, and R.~M. {Gonz{\'a}lez
  Delgado}, \emph{\apj} \textbf{593}, 127--141 (2003).

\bibitem[{Jim{\'e}nez-Bail{\'o}n} et~al.(2005)]{jb05b}
E.~{Jim{\'e}nez-Bail{\'o}n}, M.~{Santos-Lle{\'o}}, M.~{Dahlem}, M.~{Ehle},
  J.~M. {Mas-Hesse}, M.~{Guainazzi}, T.~M. {Heckman}, and K.~A. {Weaver},
  \emph{\aap} \textbf{442}, 861--877 (2005).

\bibitem[{Levenson} et~al.(2005)]{levenson05}
N.~A. {Levenson}, K.~A. {Weaver}, T.~M. {Heckman}, H.~{Awaki}, and
  Y.~{Terashima}, \emph{\apj} \textbf{618}, 167--177 (2005).

\bibitem[{Cid Fernandes} et~al.(2001)]{cidfernandes01}
R.~{Cid Fernandes}, T.~{Heckman}, H.~{Schmitt}, R.~M.~G. {Delgado}, and
  T.~{Storchi-Bergmann}, \emph{\apj} \textbf{558}, 81--108 (2001).

\bibitem[{Blustin} et~al.(2005)]{blust05}
A.~J. {Blustin}, M.~J. {Page}, S.~V. {Fuerst}, G.~{Branduardi-Raymont}, and
  C.~E. {Ashton}, \emph{\aap} \textbf{431}, 111--125 (2005).

\bibitem[{Elvis}(2000)]{elvis00}
M.~{Elvis}, \emph{\apj} \textbf{545}, 63--76 (2000).

\bibitem[{Nagao} et~al.(2006)]{nagao06}
T.~{Nagao}, R.~{Maiolino}, and A.~{Marconi}, \emph{\aap} \textbf{447}, 863--876
  (2006).

\end{thebibliography}

\IfFileExists{\jobname.bbl}{}
 {\typeout{}
  \typeout{******************************************}
  \typeout{** Please run "bibtex \jobname" to optain}
  \typeout{** the bibliography and then re-run LaTeX}
  \typeout{** twice to fix the references!}
  \typeout{******************************************}
  \typeout{}
 }

\end{document}